\documentclass[aps, prl,epsfigure,twocolumn, superscriptaddress
]{revtex4-1}
\usepackage{graphicx}
\usepackage{amsmath}    
\usepackage{amsfonts}
\usepackage[colorlinks, citecolor=red]{hyperref}
\usepackage{hyperref}
\usepackage{color}
\usepackage{epsfig}
\usepackage[up]{subfigure}
\usepackage{epstopdf}
\newcommand{\be}{\begin{equation}}
\newcommand{\ee}{\end{equation}}
\newcommand{\bc}{\begin{center}}
\newcommand{\ec}{\end{center}}
\newcommand{\bea}{\begin{eqnarray}}
\newcommand{\eea}{\end{eqnarray}}
\newcommand{\ba}{\begin{array}}
\newcommand{\ea}{\end{array}}

\def\bra#1{\mathinner{\langle{#1}|}}
\def\ket#1{\mathinner{|{#1}\rangle}}

\newcommand{\miniket}[1]{\vert#1\rangle}

\newcommand{\miniprod}[2]{\langle#1\vert#2\rangle}
\newcommand{\sand}[3]{\langle#1\vert#2\vert#3\rangle}

\newtheorem{theo}{Theorem}

\newenvironment{proof}[1][Proof]{\begin{trivlist}
\item[\hskip \labelsep {\bfseries #1}]}{\end{trivlist}}
\newcommand{\qed}{\nobreak \ifvmode \relax \else
      \ifdim\lastskip<1.5em \hskip-\lastskip
      \hskip1.5em plus0em minus0.5em \fi \nobreak
      \vrule height0.75em width0.5em depth0.25em\fi}

\begin{document}

\title{Self-avoiding quantum walks: realisations in subspaces and limit theorems}

\author{T.~Machida}
\email{machida@stat.t.u-tokyo.ac.jp}
\affiliation{Japan Society for the Promotion of Science, Japan}
\affiliation{Department of Mathematics, University of California, Berkeley, USA}
\author{C. M.~Chandrashekar}
\email{c.madaiah@oist.jp}
\affiliation{Quantum Systems Unit, Okinawa Institute of Science and Technology Graduate University, Okinawa, Japan}
\author{N.~Konno}
\email{konno@ynu.ac.jp}
\affiliation{Department of Applied Mathematics, Faculty of Engineering,
Yokohama National University, Hodogaya, Yokohama, 240-8501, Japan}
\author{Th.~Busch}
\email{thomas.busch@oist.jp}
\affiliation{Quantum Systems Unit, Okinawa Institute of Science and Technology Graduate University, Okinawa, Japan}


\begin{abstract}
While completely self-avoiding quantum walks have the distinct property of leading to a trivial unidirectional transport of a quantum state, an interesting and non-trivial dynamics can be constructed by restricting the self-avoidance to a subspace of the complete Hilbert space.  Here, we present a comprehensive study of three two-dimensional quantum walks, which are self-avoiding in coin space, in position space and in both, coin and position space. We discuss the properties of these walks and show that all result in delocalisation of the probability distribution for initial states which are strongly localised for a walk with a standard Grover coin operation. We also present analytical results for the evolution in the form of limit distributions for the self-avoiding walks in coin space and in both, coin and position space.
\end{abstract}

\maketitle

{\it Introduction~-~} Self-avoiding random walks (SRW) describe a classical walker moving on a lattice under a condition that forbids to re-visit any site that has been previously occupied\,\cite{MS93}. Such a model describes many physical or biological processes where, for example, the volume of chain-like entities restricts multiple occupancy of the same spatial positions. Folding of polymers is one such process\,\cite{MS93, Hug95}. Though extensive numerical work on SRW has led to many interesting and useful insights, not many results are known analytically\,\cite{CG96, GC01}. 

The quantum mechanical analogue to the random walk is the so called discrete-time quantum walk\,\cite{Ria58+, ADZ93, ABN01+, Kon02}, which evolves a single quantum state using discrete steps on a discrete lattice in position space. The total quantum state for the walk is described on the tensor Hilbert space $\mathcal{H}_p\otimes\mathcal{H}_c$, where $\mathcal{H}_p$ and $\mathcal{H}_c$ are the position and coin Hilbert space spanned by the position basis states and internal states of the walker, respectively. One of its main features is the fact that it creates a coherent superposition of the initial state at distinct lattice sites, which leads to an evolution through multiple path and therefore to defining interference effects. These interference effects result in a quadratically faster spread of the probability distribution compared to the classical random walk\,\cite{ABN01+}. Since a direct quantum analog of the self-avoiding walk would forbid a quantum walker to revisit the positions it previously occupied, all interference effects would be suppressed and a trivial, unidirectional transport of the basis states of the walker would result. For this reason completely self-avoiding quantum walk (SQW) have not created much interest, though quantum walks have been studied extensively for over a decade now.

 Only recently a non-trivial form of a SQW, which forbids the quantum state of the walker to evolve onto itself in coin space, was introduced and studied numerically\,\cite{BPH13}.  The basic idea is that self-avoidance is possible in the subspace of the points of the position space which form the outer edge of the distribution. In these positions the walker is only in one of the basis states and a coin operation that evolves only onto the orthogonal basis states can be constructed. Inside the distribution the walker is in a superposition of all basis states and self-avoidance is not possible. While for a two-state walker in one dimension this form of a SQW in the coin space still leads to a suppressing of interference and therefore trivial and unidirectional transport of the quantum state,  Barr {\sl et al.} \cite{BPH13} showed that considering a four-state walker on a two-dimensional lattice results in interesting and nontrivial dynamics.

Here we significantly extend the class of SQW by introduce two new walks, one of them self-avoiding in position space and one self-avoiding in both, coin and position space, and compare these to the known case.  Interestingly, we find that these self-avoiding walks result in delocalisation of the probability distribution for initial states that show strong localisation for walks using the standard Grover coin operations. To gain analytical insight into this behaviour, we derive the associated forms of the limit distributions for the SQW in coin space and for the SQW in both, coin and position space and compare them to the well-known limiting function for the Grover walk. Consistently, the most striking feature is the absence of the Dirac $\delta-$function at the origin for both distributions, which is responsible for the the localisation in the Grover walk and confirms the delocalising properties of the SQWs. While such a feature might not be unexpected for a self-avoiding process, we show that the detailed form differs strongly for each walk. 

In this work we focus mainly on the mathematical treatment of the walks, however their applications to physical systems are interesting to consider as well. They can, for example, be used to describe the behaviour of quantum dimers as suggested by Barr\,{\em et al.}~\cite{BPH13} or give insights into interacting two- and many particle quantum walks\,\cite{AMS12, LVH12, CB12}. In these the return to  previous positions in coin or position space can be restricted due to interaction with the other particles. As these SQW are defined using a new set of coin operations, they can also be considered for new applications in quantum information processing.



\begin{figure}[ht]
\bc 
\subfigure[ $\alpha = -\beta =-\gamma = \delta = \frac{1}{2}$]{\includegraphics[width=4.4cm]{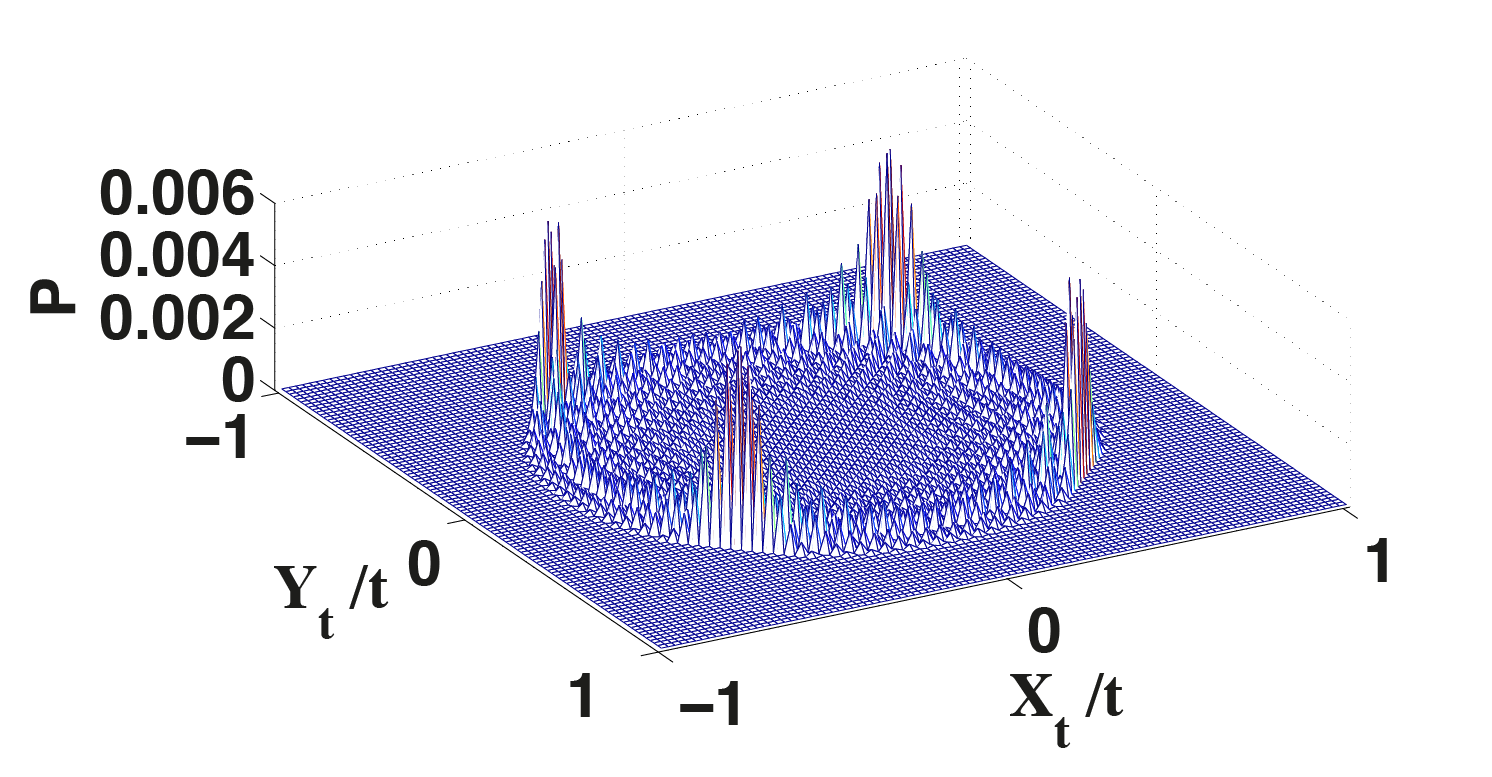} 
\label{fig:1a}}
\hskip -0.21in 
\subfigure[  $\alpha = -\delta = \frac{1}{2}$, $\beta =\gamma = \frac{i}{2}$]{\includegraphics[width=4.4cm]{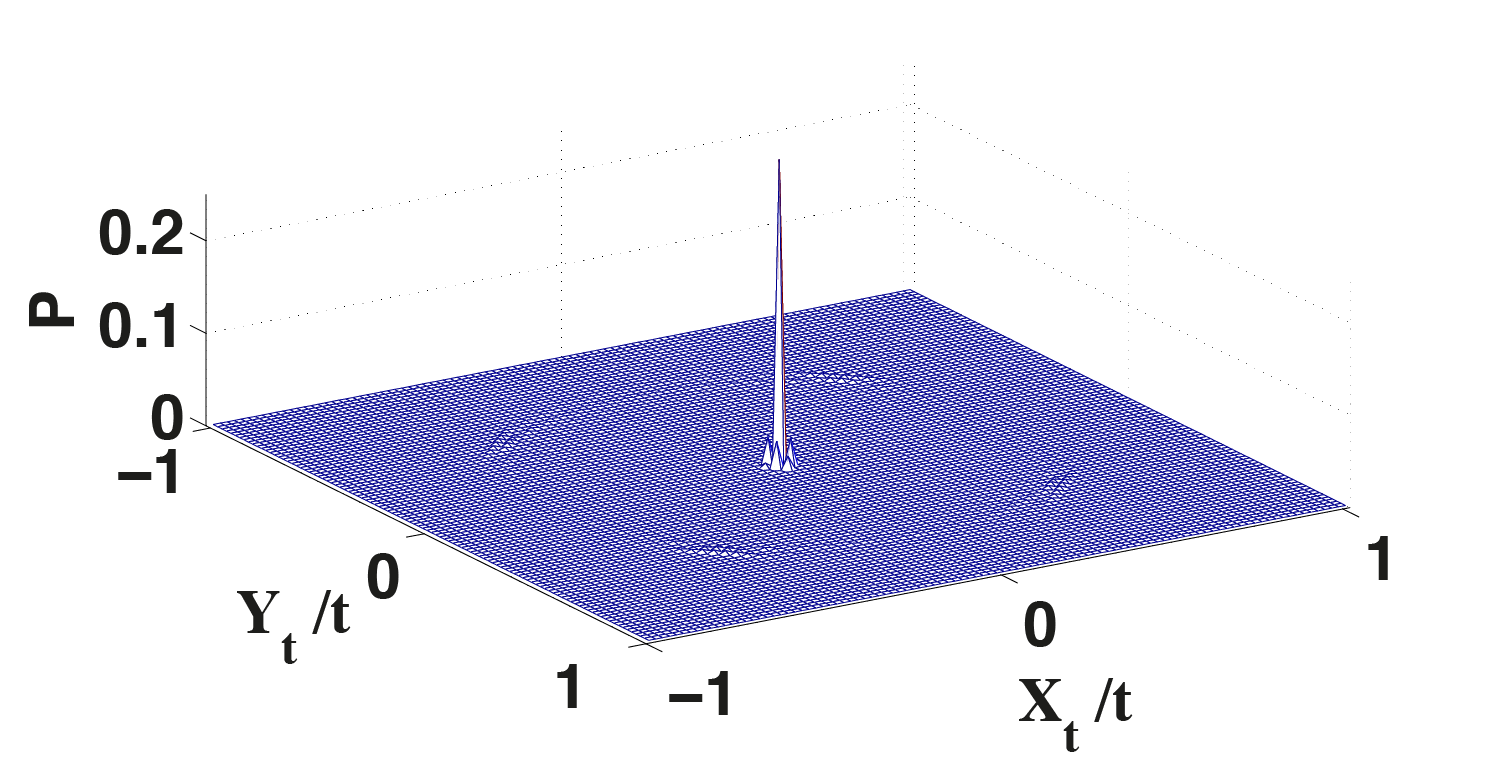} 
\label{fig:1b}}
\ec
\caption{\footnotesize{(Color online)  Probability distribution $P(X_t/t,Y_t/t)$ of the Grover walk with different initial states after $t=100$ steps. In (a) the initial state is characterised by $\alpha = -\beta =-\gamma = \delta = \frac{1}{2}$ and a wide spread probability distribution is clearly visible, whereas in (b) an example for a strongly localised distribution is shown stemming from an initial state with $\alpha = -\delta = \frac{1}{2}$, $\beta =\gamma = \frac{i}{2}$. \label{fig:1}}}
\end{figure}

{\it Quantum walk on a two-dimensional lattice~-~} The standard example for a quantum walk on a two-dimensional lattice is the so called Grover walk\,\cite{MBS02, TFM03, IKK04, WKK08}.  For a Grover walk, $\mathcal{H}_p$ is a two-dimensional position space spanned by the basis $\left\{\miniket{x,y} : x,y\in\mathbb{Z}\right\}$ and $\mathcal{H}_c$ is a coin space spanned by the basis $\left\{\miniket{l}, \miniket{u}, \miniket{d}, \miniket{r}\right\}$
\begin{equation}
 \ket{l}=\left[\begin{array}{c}
	  1\\0\\0\\0
	       \end{array}\right],
 \ket{u}=\left[\begin{array}{c}
	  0\\1\\0\\0
	       \end{array}\right],
 \ket{d}=\left[\begin{array}{c}
	  0\\0\\1\\0
	       \end{array}\right],
 \ket{r}=\left[\begin{array}{c}
	  0\\0\\0\\1
	       \end{array}\right].
\end{equation}
The state $\ket{\Psi_t}\,\in\mathcal{H}_p\otimes\mathcal{H}_c$ at time $t\,\in\left\{0,1,2,\ldots\right\}$ evolves to the next state $\ket{\Psi_{t+1}}$ by first applying a coin-flip operator $C$, followed by a position-shift operator $S$, so that
\begin{equation}
 \miniket{\Psi_{t}}=(SC)^t\miniket{\Psi_0},
\end{equation}
where 
\begin{align}
\label{ins}
\miniket {\Psi_0} = \ket{0,0}\otimes\Bigl ( \alpha\miniket{l} + \beta\miniket{u} + \gamma\miniket{d} + \delta\miniket{r} \Bigr ),
\end{align}
is the initial state at the origin with $\alpha,\beta,\gamma$ and $\delta$ being complex numbers satisfying the condition $|\alpha|^2+|\beta|^2+|\gamma|^2+|\delta|^2=1$.
The coin and the position-shift operators are given by 
\bea
    C= \sum_{x,y\in\mathbb{Z}}\ket{x,y}\bra{x,y}\otimes \frac{1}{2}\left[\begin{array}{cccc}
		     -1  &~~1&~~1& ~~1\\
			   ~~1&-1&~~1&~~1\\
			   ~~1&~~1&-1&~~1\\
			   ~~1&~~1&~~1&-1
			  \end{array}\right], \label{eq:grovercoin}
			  \eea
and			  
	\bea		  
	\label{shift1}
	S=\sum_{x,y\in\mathbb{Z}}\Bigl(\ket{x-1,y}\bra{x,y}\otimes\ket{l}\bra{l}  
        + \ket{x,y+1}\bra{x,y}\otimes\ket{u}\bra{u} \nonumber \\
       +\ket{x,y-1}\bra{x,y}\otimes\ket{d}\bra{d}
       +\ket{x+1,y}\bra{x,y}\otimes\ket{r}\bra{r}\Bigr).~~~~
       \label{eq:shift-operator}
 \eea
The probability distribution of the walker at time $t$ is then given by,  
\begin{align}
 &P(X_t,Y_t) 
 =\bra{\Psi_t}\left(\ket{x,y}\bra{x,y}\otimes\sum_{j} \ket{j}\bra{j}\right)\ket{\Psi_t},
\end{align}
where $(X_t,Y_t)$ denotes the position $(x, y)$ of walker and $j\in\left\{l,u,d,r\right\}$.  It is shown for $t=100$ steps and for an initial state $\alpha = -\beta =-\gamma = \delta = \frac{1}{2}$ in Fig.\,\ref{fig:1a}, as a function of the scaled variables $X_t/t$ and $Y_t/t$. The wide spread in position space for this specific initial state is clearly visible\,\cite{TFM03, IKK04}. However, all other initial states are known to lead to strong localisation around the initial position for the Grover coin and an example for an initial state with $\alpha = -\delta = \frac{1}{2}$, $\beta =\gamma = \frac{i}{2}$ is shown in Fig.\,\ref{fig:1b}.  

\begin{figure}[ht]
\bc 
\subfigure[~Self-avoiding in coin space]{\includegraphics[width=6.0cm]{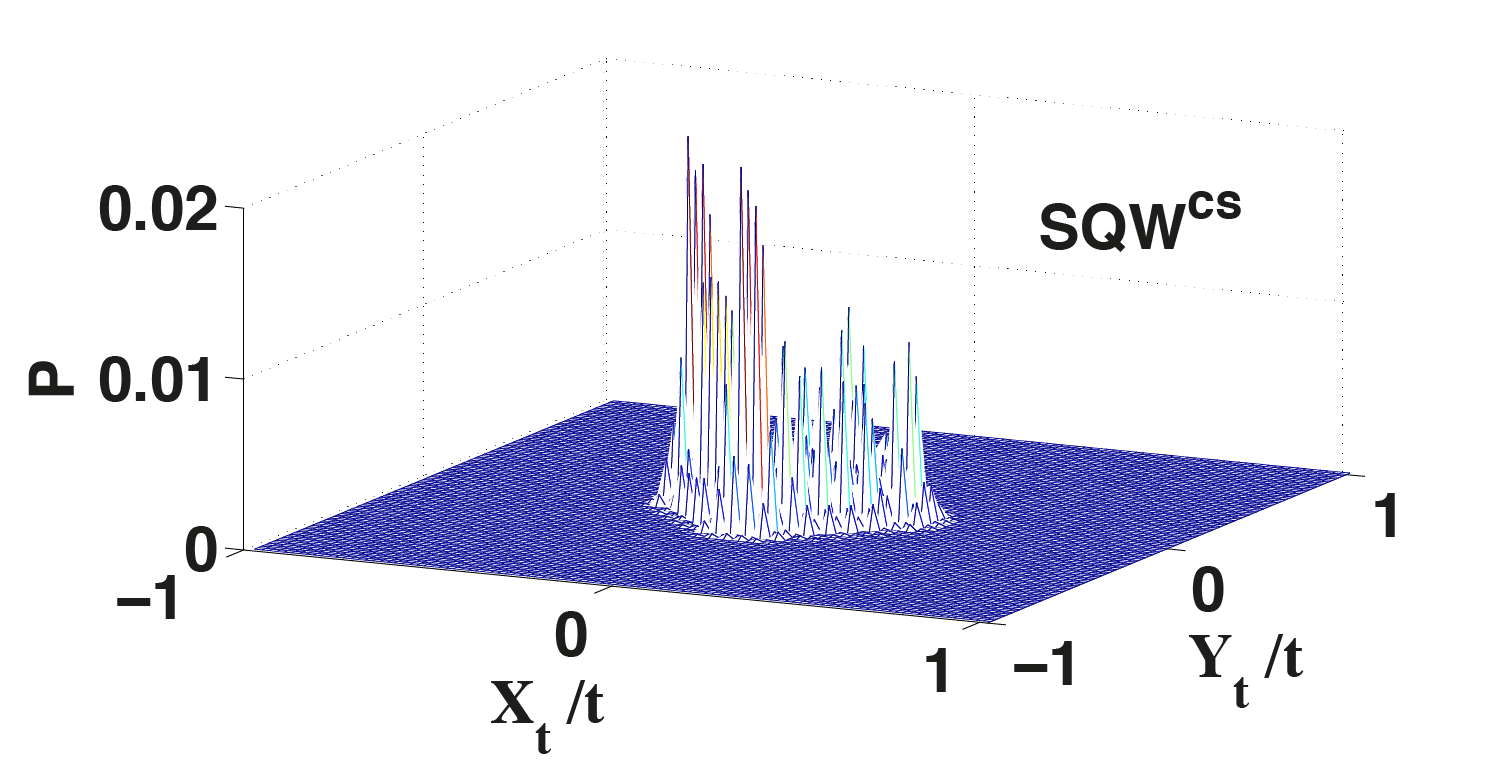} 
\label{fig:2a}}
\subfigure[~Self-avoiding in position space]{\includegraphics[width=6.0cm]{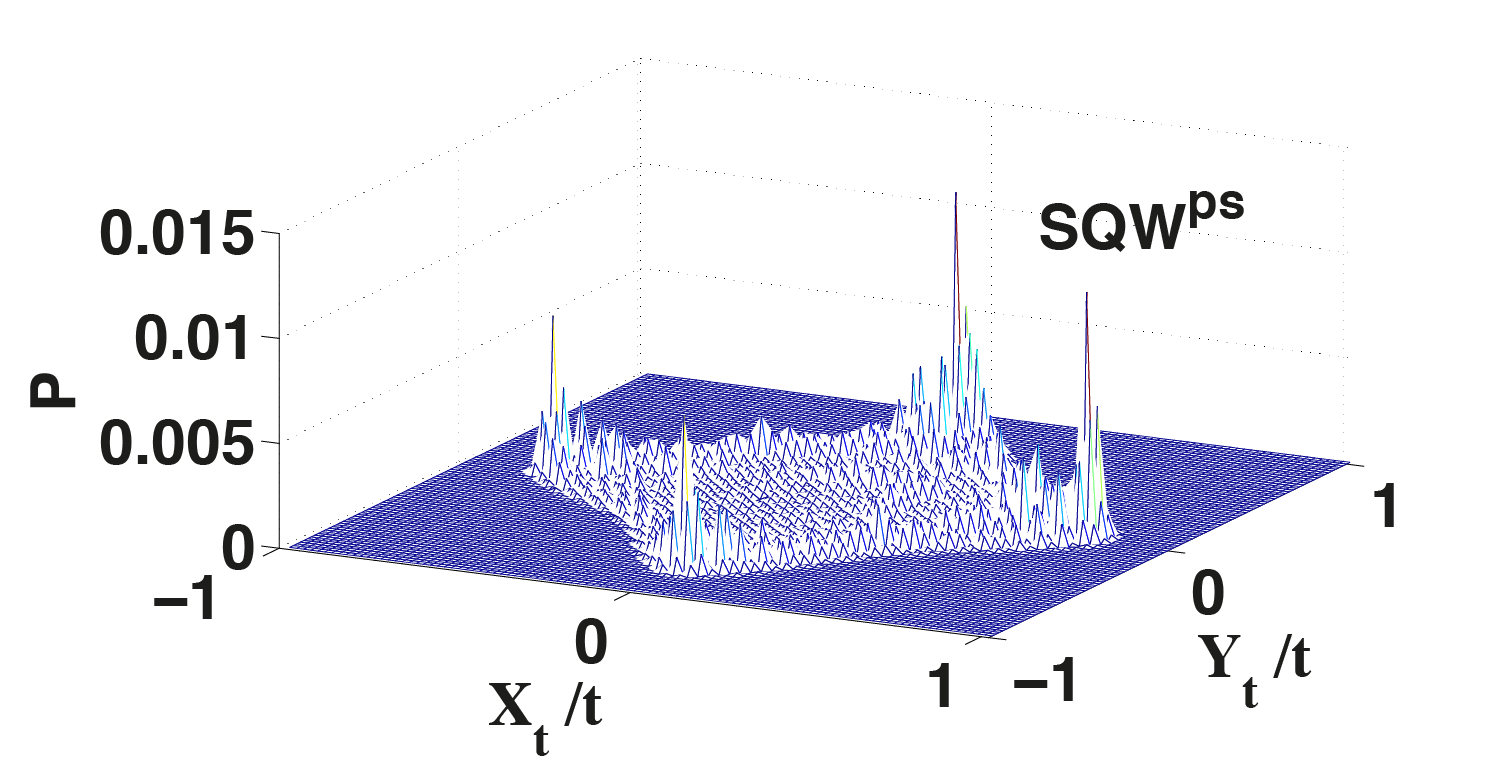} 
\label{fig:2b}}
\subfigure[~Self-avoiding in coin and position space]{\includegraphics[width=6.0cm]{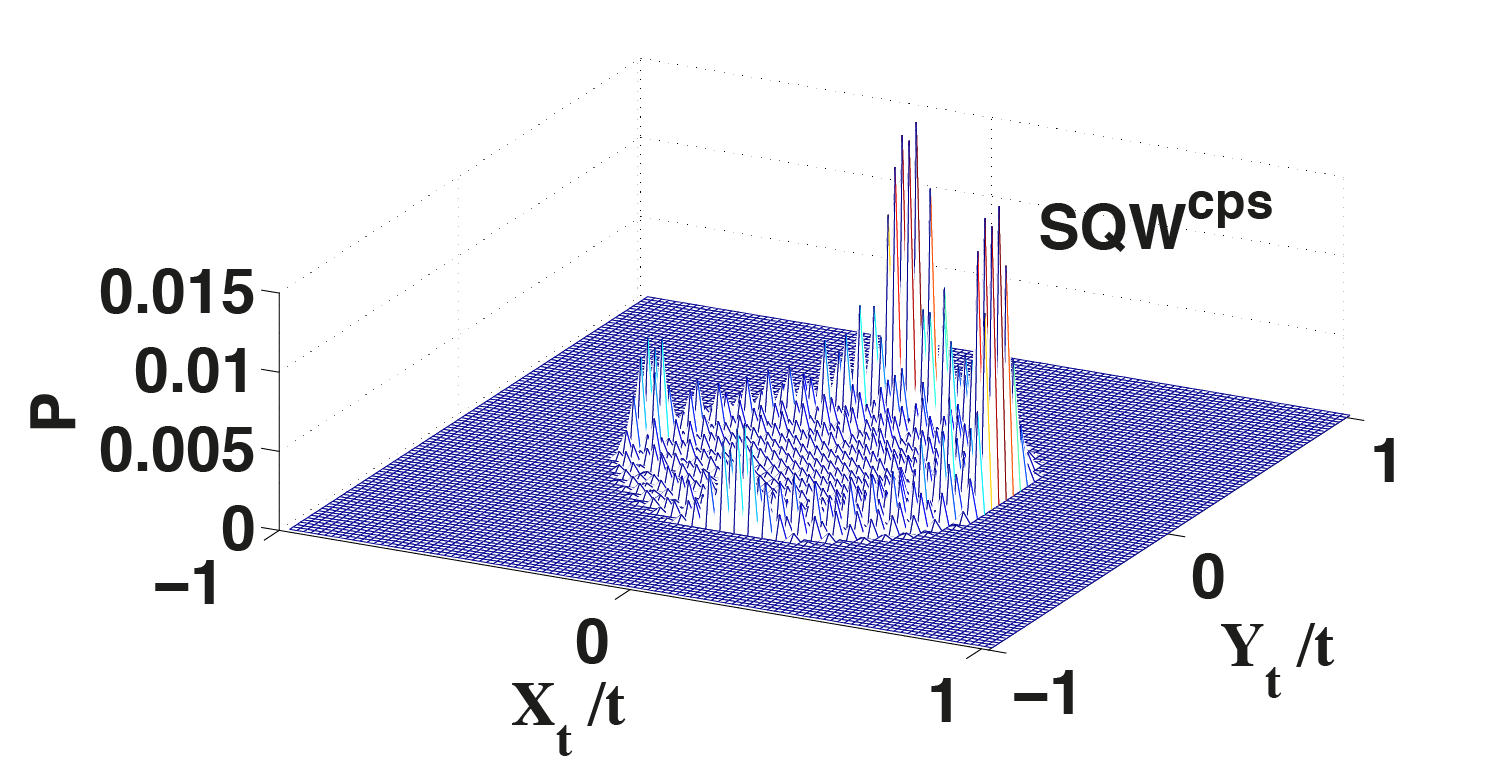} 
\label{fig:2c}}
\ec
\caption{\footnotesize{(Color online) Probability distribution of the SQW in subspace after $t=100$ for the walker with $\alpha = -\delta = \frac{1}{2}$, $\beta =\gamma = \frac{i}{2}$ in the initial state. For the Grover walk with the same initial state we get a localised distribution but for all the three forms of SQWs in the subspace we obtain a distinct delocalised probability distribution.
\label{fig:2}}}
\end{figure}

{\it Self-avoiding in coin space~-~} For comparison and as a reminder, let us briefly review the self-avoiding walk in coin space \cite{BPH13}, where the walker in any particular basis state $|j\rangle$ $(j \in \{l, u, d, r\})$ is only allowed to evolve into a superposition of the basis state other than itself. For this all $|j\rangle\langle j|$-components in coin operation have to be zero, which leads to the coin operator\,\cite{BPH13},
 \bea
 \label{sqwcs}
 C^{sc}=
 \sum_{x,y\in\mathbb{Z}}\ket{x,y}\bra{x,y}\otimes\frac{1}{\sqrt{3}}\left[\begin{array}{cccc}
		     ~~0&~~1&~~1&-1\\
			   ~~1&~~0&~~1&~~1\\
			   ~~1&-1&~~0&-1\\
			   -1&-1&~~1&~~0
			  \end{array}\right].
\eea
During the evolution of the walk, apart from the positions at the edge of the evolving position space, the walker will be in a superposition of more than one of the basis   states. Therefore, the coin operation given by Eq.\,(\ref{sqwcs}), will not be able to block any of the basis states, except at the positions on the outer edge of the evolving position space. This self-avoiding in the subspace of the outer edge of the evolving position space, however, does significantly affect the interference pattern and results in a deviation of the probability distribution when compared to the standard Grover evolution. To demonstrate this significance, we show in Fig.\,\ref{fig:2a} the probability distribution 
for the SQW in coin space with $\alpha = -\delta = \frac{1}{2}$, $\beta =\gamma = \frac{i}{2}$ as the initial state. While a Grover walk with the same initial state results in a localised distribution (see Fig.\,\ref{fig:2b}), the SQW delocalises the probability distribution.

{\it Self-avoiding in position space~-~} A walk that is self-avoiding in position space can be constructed by ensuring that the walker in any particular position at time $t$ does not revisit the same position it occupied at $t-1$. To achieve this, the coin operation must be chosen such that a state $\ket{l}$ (resp. $\ket{u}, \ket{d}$ or $\ket{r}$)  present at position $|x, y\rangle$ at time $t-1$ does not flip to $\ket{r}$ (resp. $\ket{d}, \ket{u}$ or $\ket{l}$), which requires taking the $|r\rangle \langle l | , |d\rangle \langle u | , |u \rangle \langle d | , |l\rangle \langle r | $ -components in the coin operation to be zero (see position-shift operator Eq.~(\ref{shift1})). This leads to the coin-flip operator  
 \bea
 C^{sp}=
 \sum_{x,y\in\mathbb{Z}}\ket{x,y}\bra{x,y}\otimes\frac{1}{\sqrt{3}}\left[\begin{array}{cccc}
		     -1&-1&~~1&~~0\\
		     ~~1&-1&~~0&-1\\
		   ~~1&~~0&~~1&~~1\\
		     ~~0&~~1&~~1&-1	   
			  \end{array}\right].
\eea
In the same way as discussed for the SQW in coin space, the evolution for SQW in position space will also result in self-avoiding only at the outer edges of the distribution in position space.  The probability distribution  after $t=100$ steps of evolution is shown in Fig.~\ref{fig:2b} for the walker with an initial state characterised by $\alpha = -\delta = \frac{1}{2}$, $\beta =\gamma = \frac{i}{2}$  and again a delocalised distribution, a significant change to the Grover walk, but also to the SQW in coin space, is clearly visible.

{\it Self-avoiding in coin and position space~-~} Combining both restrictions described above leads to a SQW in both, coin and position space. In this walk the state is not allowed to evolve onto itself and the walker cannot  revisit any position it currently occupies. To achieve this, the coin operation is chosen such that both restrictions are fulfilled simultaneously, i.e.~each state $\ket{j}\,(j\in\left\{l,u,d,r\right\})$ does not flip onto itself and the state $\ket{l}$ (resp. $\ket{u}, \ket{d}$ or $\ket{r}$)  which was at position $|x, y\rangle$ a step before 
does not evolve onto $\ket{r}$ (resp. $\ket{d}, \ket{u}$ or $\ket{l}$). This leads to a  coin-flip operator of the form 
\bea
 C^{scp}=
 \sum_{x,y\in\mathbb{Z}}\ket{x,y}\bra{x,y}\otimes\frac{1}{\sqrt{2}}\left[\begin{array}{cccc}
		     ~~0 &-1&~~1&~~0\\
		     ~~1&~~0&~~0&-1\\
		   ~~1&~~0&~~0&~~1\\
		     ~~0&~~1&~~1&~~0	   
			  \end{array}\right].
\eea
and in Fig.\,\ref{fig:2c} we show the probability distribution for this case after $t=100$ steps for the same initial state as before ($\alpha = -\delta = \frac{1}{2}$, $\beta =\gamma = \frac{i}{2}$). Again a delocalised distribution is obtained, with a spread between the one achieved for the SQW in coin space and the one in position space. This case also shows the most distinctive asymmetry of the three discussed. 

We have now seen that all three SQW presented above lead to a delocalised distribution for an initial state defined by 
$\alpha = -\delta = \frac{1}{2}$, $\beta =\gamma = \frac{i}{2}$, which has a localised distribution for the Grover walk.  While the self-avoidance in position space results in a larger spread compared to the other two SQW's, the self-avoidance in coin space discourages the outward spread at the edge of the evolution and therefore results in a smaller distribution. This delocalisation is a general feature which holds true for any other initial state which localise for pure Grover walk dynamics, and can be explained by the nature of self-avoidance. Therefore these three coin operations are in the same class of operations as the ones presented in Ref.\,\cite{IKK04} which also delocalise the distribution.

{\it Limit Theorem~-~} While the numerical simulations above give an impression of the short term behaviour of the walks, in the following we present the long-time limit distributions for the two-dimensional SQW in coin space\,\cite{BPH13} and in both, coin and position space. For reference, however, let us briefly review the long-time limit distribution of the Grover walk, which was obtained by Watabe {\sl et al.}\,\cite{WKK08} as
\bea
   \lim_{t\to\infty} P&&\left(\frac{X_t}{t}\leq x,\,\frac{Y_t}{t}\leq y\right)\nonumber\\
 =&&\int_{-\infty}^x du\,\int_{-\infty}^y dv\,\biggl\{\Delta(\alpha,\beta,\gamma,\delta)\,\delta_o(u,v)\nonumber\\
 &&\;\;\qquad+f(u,v)\eta(u,v;\alpha,\beta,\gamma,\delta)I_{\mathcal{D}}(u,v)\biggr\}.
\eea
Here $\delta_o(x,y)$ is the Dirac $\delta$-function at the origin and the first part of the integral describes the localised part of the distribution. Note that its pre-factor 
 $\Delta(\alpha,\beta,\gamma,\delta)=\frac{1}{\pi}\Re\Big[(\pi -2)(\alpha+\delta)(\bar\beta+\bar\gamma)  +(\pi-4) (\alpha\bar\delta+\beta\bar\gamma) \Big] + \frac{1}{2}$,
goes to zero for the specific initial state with $\alpha=-\beta=-\gamma=\delta=\frac{1}{2}$. The non-localised part is given by $f(x,y)$, $\eta(x,y;\alpha,\beta,\gamma,\delta)$, and $I_{\mathcal{D}}(x,y)$ and the exact definition of these terms can be found in \,\cite{supp}.

\begin{figure}[ht]
\bc 
\subfigure[~Self-avoiding in coin space]{\includegraphics[width=4.1cm]{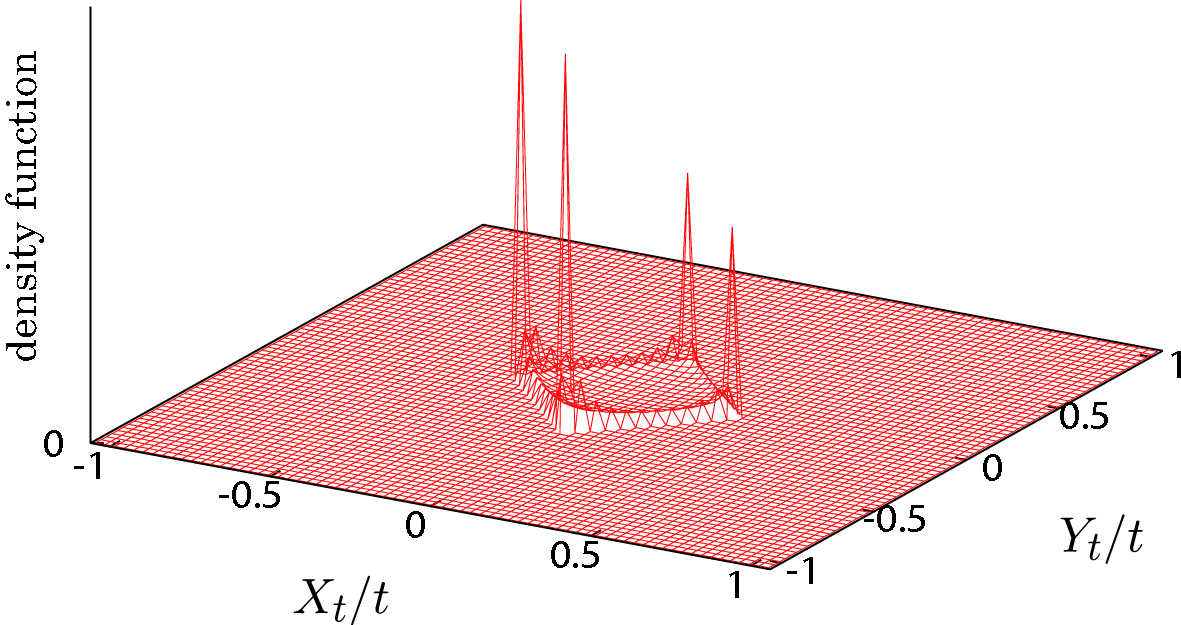} 
\label{fig:3a}}
\subfigure[~Self-avoiding walk in coin and position space]{\includegraphics[width=4.1cm]{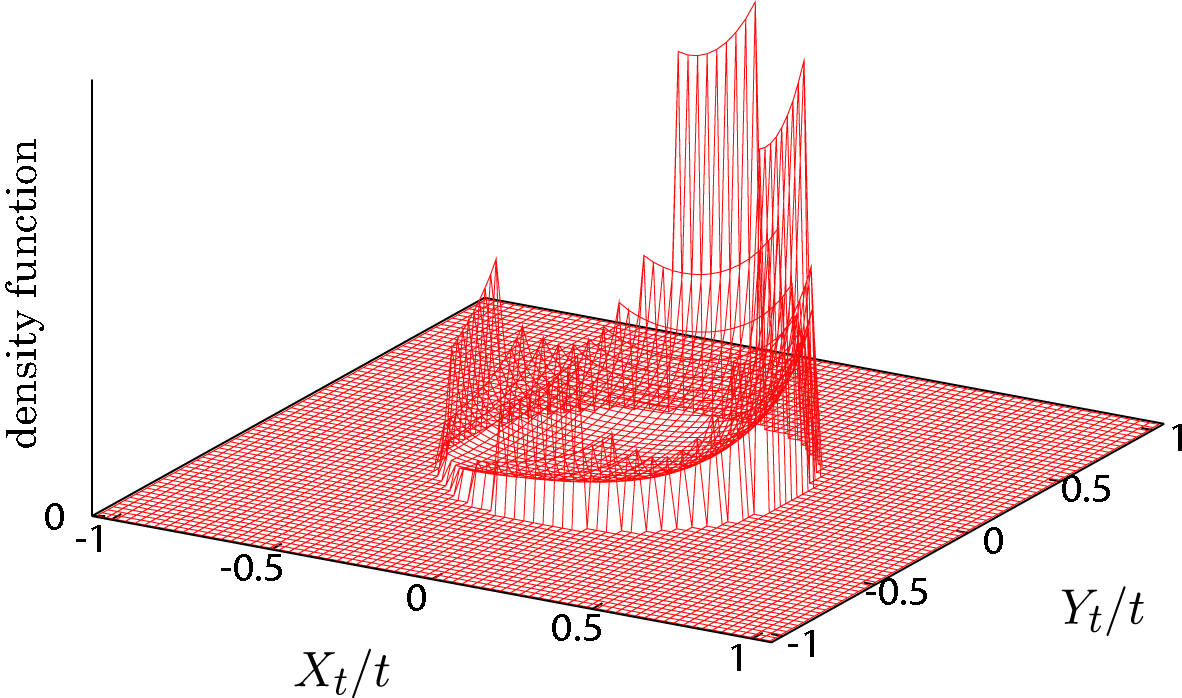} 
\label{fig:3b}}
\ec
\caption{\footnotesize{(Color online) The limit density function as $t\to\infty$ for walker with $\alpha=-\delta=\frac{1}{2},\, \beta=\gamma=\frac{i}{2}$ in the initial state 
for (a) self-avoiding in  coin space and (b) for self-avoiding in coin and position space. Both limit density functions reproduce the main features of the respective probability distributions from the discrete-time evolution.
\label{fig:3}}}
\end{figure}

\begin{theo}
For the SQW in coin space governed by the coin operator $C^{sc}$, the long-time limit distribution is given by
\begin{align}
 \lim_{t\to\infty}P&\left(\frac{X_t}{t}\leq x,\,\frac{Y_t}{t}\leq y\right)\nonumber\\
 =&\int_{-\infty}^x du\,\int_{-\infty}^y dv\,f_{sc}(u,v)\left\{g_1(u,v)+g_2(u,v)\right\}\nonumber\\
 &\times\eta_{sc}(u,v;\alpha,\beta,\gamma,\delta)I_{\mathcal{D}_{sc}}(u,v),
\end{align}
where
\begin{align}
 f_{sc}(x,y)=&\frac{1}{\pi^2(1-4x^2)(1-4y^2)\sqrt{D_{sc}(x,y)}}, \nonumber \\
 D_{sc}(x,y)=&81(x^4+y^4)-18x^2y^2-18(x^2+y^2)+1,\nonumber \\
 g_1(x,y)=&\Bigl\{648(x^4+y^4)+576x^2y^2-324(x^2+y^2)\nonumber\\
 &+53+4\left\{7-18(x^2+y^2)\right\}\sqrt{D_{sc}(x,y)}\Bigr\}^{\frac{1}{2}}, \nonumber \\
 g_2(x,y)=&\Bigl\{648(x^4+y^4)+576x^2y^2-324(x^2+y^2)\nonumber \\
 &+53-4\left\{7-18(x^2+y^2)\right\}\sqrt{D_{sc}(x,y)}\Bigr\}^{\frac{1}{2}},\nonumber \\
 \eta_{sc}(x,y;\alpha&,\beta,\gamma,\delta) =1+2x\Bigl[|\alpha|^2-|\delta|^2\nonumber\\
     &\quad+\Re\left\{(-\alpha+\gamma-\delta)\overline{\beta}+(\alpha+\beta-\delta)\overline{\gamma}\right\}\Bigr]\nonumber\\
 -2y\Bigl[|\beta|^2 & -|\gamma|^2 +\Re\left\{(\beta+\gamma+\delta)\overline{\alpha}+(\alpha-\beta+\gamma)\overline{\delta}\right\}\Bigr], \nonumber \\
  I_{\mathcal{D}_{sc}}(x,y)=&\left\{\begin{array}{ll}
			 1&\left(D_{sc}(|x|,|y|)>0,\,0\leq |x|,|y|\leq \frac{1}{3}\right),\\[1mm]
			       0&(\mbox{otherwise}).\end{array}\right. \nonumber 
\end{align}
\end{theo}
The proof is given in the supplementary material \,\cite{supp} and to visualise this result we show in Fig.~\ref{fig:3a} the limit density function, which should be compared with the finite time one shown in Fig.~\ref{fig:2a}.  Very similar features are clearly visible and in particular it also shows the expected delocalisation. Since the distribution shown in Fig.\,\ref{fig:2a} captures the state of the walker at $t=100$, not all features are as clearly developed as in the limit density function, however the overall form and asymmetry are very similar.

\begin{theo}
For the SQW in both, coin and position space determined by the coin operation $C^{scp}$, we have the following long-time limit distribution.
\begin{align}
 &\lim_{t\to\infty} P\left(\frac{X_t}{t}\leq x,\,\frac{Y_t}{t}\leq y\right)\nonumber\\
 =&\int_{-\infty}^x du\,\int_{-\infty}^y dv\,f_{scp}(u,v)\eta_{scp}(u,v;\alpha,\beta,\gamma,\delta)I_{\mathcal{D}_{scp}}(u,v),
\end{align}
where
\begin{align}
 f_{scp}(x,y)=&\frac{4}{\pi^2(1-4x^2)(1-4y^2)}, \nonumber \\[5mm]
 \eta_{scp}(x,y;\alpha,\beta,\gamma,\delta)
 =&1-2x\left\{|\alpha|^2-|\delta|^2-2\Re(\beta\bar{\gamma})\right\}\nonumber\\
 &+2y\left\{|\beta|^2-|\gamma|^2-2\Re(\alpha\bar{\delta})\right\}, \nonumber \\[5mm]
 I_{\mathcal{D}_{scp}}(x,y)=&\left\{\begin{array}{ll}
			 1&\left(x^2+y^2<\frac{1}{4}\right),\\[1mm]
			       0&(\mbox{otherwise}).\end{array}\right. \nonumber
\end{align}
\end{theo}
This theorem can be proven in a manner very similar to the one used for Theorem~1 \,\cite{supp} and we show the distribution
 in Fig.~\ref{fig:3b}. Again one can see that the limit density function reproduces the characteristic features of the probability distribution from the discrete-time evolution (see Fig.~\ref{fig:2c}).
\par
Furthermore, both theorems show that the compact support of both limit functions is different. For the  SQW in coin space it is described by a quartic dependence on the spatial variables and for the SQW in both, coin and position space it is given by a circular function. This difference is also clearly visible from Figs.\,\ref{fig:2} and \ref{fig:3}.


{\it Summary~-~} In this work we have presented two new forms of SQWs on a two-dimensional lattice, one in position space and one in position and coin space. Together with the already know form of the SQW in coin space, these constitute a new class of quantum walks with unique properties.  Unlike a SRW, which completely avoids positions previously visited, the SQW was defined such that it is self-avoiding only in a subspace of the complete Hilbert space of the system, in oder to preserve the characteristic interference properties of the quantum setting. Since the dynamics of a quantum walk can be completely controlled by the quantum coin operation we have presented two different coin operators to define the two distinct self-avoiding evolutions and shown numerical results for finite time evolutions and long-time limit theorems for a walker in rescaled coordinates $(X_t/t,Y_t/t)$. Apart from giving the explicit forms of the SQW and the associated limit theorems, one of our main findings is that the SQW in a subspace results in the delocalisation of the distribution for  initial states which show localisation for evolutions using Grover coin operation. 
This can be clearly see by the absence of the term containing a Dirac $\delta$-function at the origin in the limit distributions, which is the case for the pure Grover walk.

{\it Ackowledgements~-~}TM acknowledges support from Naoki Masuda and the Japan Society for the Promotion of Science. NK acknowledges financial support of the Grant-in-Aid for Scientific Research (C) of Japan Society for the Promotion of Science (Grant No. 21540118).



\begin{widetext}

\section{Supplementary Material}


\subsection{Details on the Limit Theorem for the Groover walk}
\label{lt}

The long-time limit distribution originating from this Grover walk was obtained by Watabe {\sl et al.}\cite{WKK08} and is given by 
\begin{equation}
   \lim_{t\to\infty}P\left(\frac{X_t}{t}\leq x,\,\frac{Y_t}{t}\leq y\right)
          =\int_{-\infty}^x du\,\int_{-\infty}^y dv\,\biggl\{\Delta(\alpha,\beta,\gamma,\delta)\,\delta_o(u,v)
            +f(u,v)\eta(u,v;\alpha,\beta,\gamma,\delta)I_{\mathcal{D}}(u,v)\biggr\},
\end{equation}
where $\delta_o(x,y)$ is the Dirac $\delta$-function at the origin and
\begin{align}
 \Delta(\alpha,\beta,\gamma,\delta)=&1-M_1-\left(\frac{1}{2}-\frac{1}{\pi}\right)(M_4+M_5),\\
 f(x,y)=&\frac{2}{\pi^2(x+y+1)(x-y+1)(x+y-1)(x-y-1)}\\
 \eta(x,y;\alpha,\beta,\gamma,\delta)=&M_1-M_2x-M_3y+M_4x^2+M_5y^2+M_6xy,
\end{align}
with
\begin{align}
 M_1=&\frac{1}{2}+\Re(\alpha\overline{\delta}+\beta\overline{\gamma}),\\[2mm]
 M_2=&|\alpha|^2-|\delta|^2+\Re(-\alpha\overline{\beta}-\alpha\overline{\gamma}+\beta\overline{\delta}+\gamma\overline{\delta}),\\[2mm]
 M_3=&-|\beta|^2+|\gamma|^2+\Re(\alpha\overline{\beta}-\alpha\overline{\gamma}+\beta\overline{\delta}-\gamma\overline{\delta}),\\[2mm]
 M_4=&\frac{1}{2}(|\alpha|^2-|\beta|^2-|\gamma|^2+|\delta|^2)
 -\Re(\alpha\overline{\beta}+\alpha\overline{\gamma}+3\alpha\overline{\delta}+\beta\overline{\gamma}+\beta\overline{\delta}+\gamma\overline{\delta}),\\[2mm]
 M_5=&-\frac{1}{2}(|\alpha|^2-|\beta|^2-|\gamma|^2+|\delta|^2)
 -\Re(\alpha\overline{\beta}+\alpha\overline{\gamma}+\alpha\overline{\delta}+3\beta\overline{\gamma}+\beta\overline{\delta}+\gamma\overline{\delta}),\\[2mm]
 M_6=&-2\Re(-\alpha\overline{\beta}+\alpha\overline{\gamma}+\beta\overline{\delta}-\gamma\overline{\delta}),\\[2mm]
 I_{\mathcal{D}}(x,y)=&\left\{\begin{array}{ll}
			 1&\left(x^2+y^2<\frac{1}{2}\right),\\[1mm]
			      0&(\mbox{otherwise}).\end{array}\right.
\end{align}
The first part of the integral describes the localised part of the distribution and one can note from the explicit form of its pre-factor 
\begin{align}
 \Delta(\alpha,\beta,\gamma,\delta)=\frac{1}{\pi}\Re\Big[(\pi -2)(\alpha+\delta)(\bar\beta+\bar\gamma)
 +(\pi-4) (\alpha\bar\delta+\beta\bar\gamma) \Big]+\frac{1}{2},
\end{align}
that it goes to zero for the specific initial state with $\alpha=-\beta=-\gamma=\delta=\frac{1}{2}$. The non-localised part is given by 
$f(x,y)$, $\eta(x,y;\alpha,\beta,\gamma,\delta)$, and $I_{\mathcal{D}}(x,y)$. Here $\Re(z)$ indicates the real part of a complex number argument.
In Fig.\ref{fig:Grover_walk}(a) the probability distribution after $t=100$ from the discrete evolution is shown and in Fig.~\ref{fig:Grover_walk}(b) the density distribution of the delocalized part of the complete limit function is shown. 
\begin{figure}
\begin{center}
 \begin{minipage}{70mm}
  \begin{center}
   \includegraphics[scale=0.6]{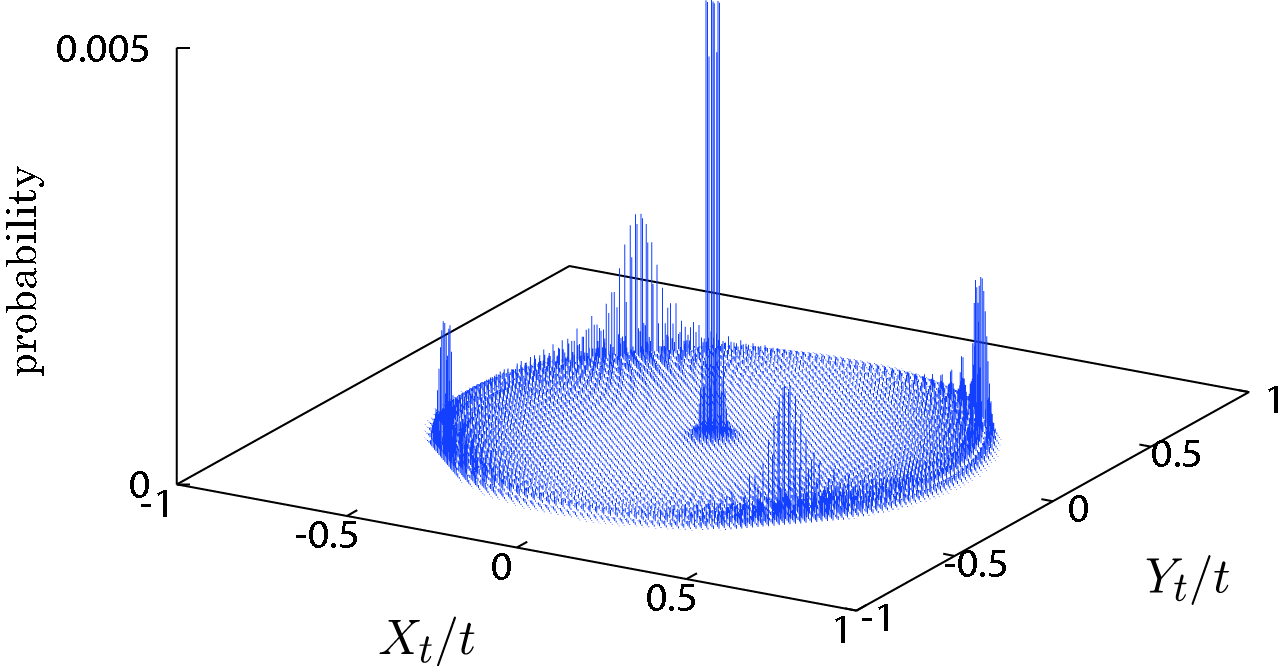}\\
   {(a) $\mathbb{P}\left[\left(\frac{X_t}{t},\frac{Y_t}{t}\right)=(x,y)\right]$}
  \end{center}
 \end{minipage}
 \vspace{10mm}
 \begin{minipage}{70mm}
  \begin{center}
   \includegraphics[scale=0.6]{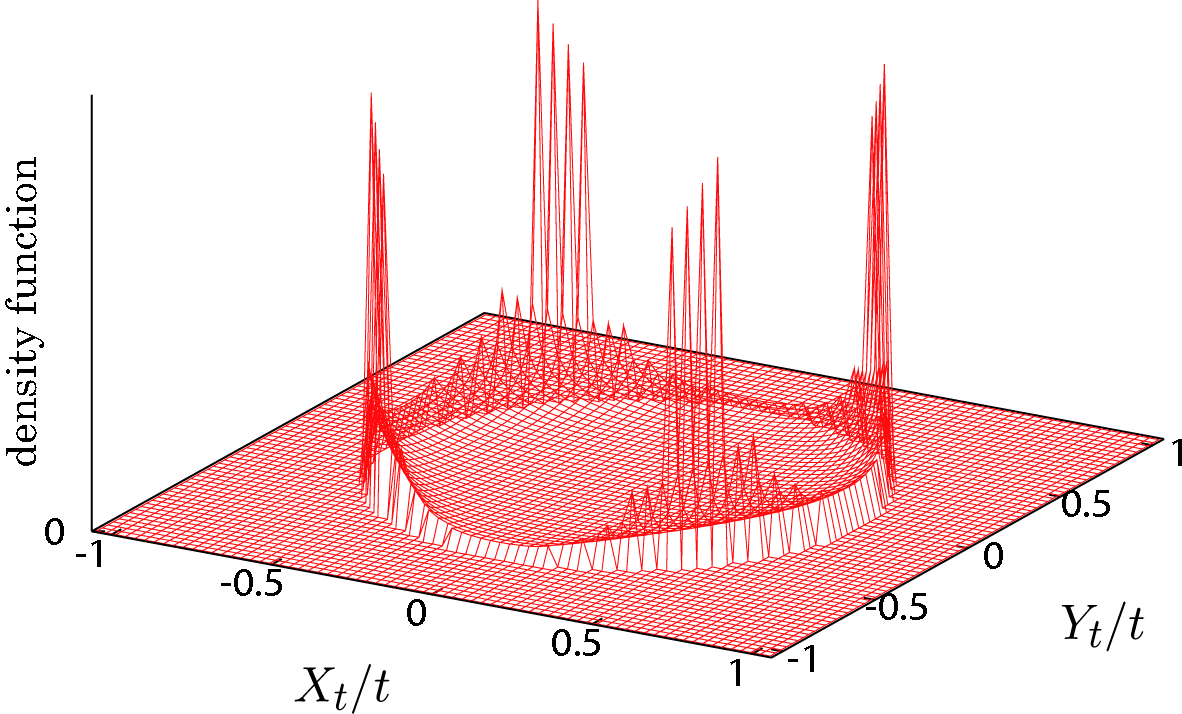}\\
   {(b) the function $f(x,y)\eta(x,y)I_{\mathcal{D}}(x,y)$}
  \end{center}
 \end{minipage}
\end{center}
\caption{Grover walk: Figure (a) shows the probability distribution $P\left[\left(\frac{X_t}{t},\frac{Y_t}{t}\right)=(x,y)\right]$ at time $t=100$. Figure (b) shows the function $f(x,y)\eta(x,y)I_{\mathcal{D}}(x,y)$. ($\alpha=-\delta=\frac{1}{2},\, \beta=\gamma=\frac{i}{2}$)}
\label{fig:Grover_walk}
\end{figure}


\subsection{Proof of the Limit theorem for self-avoiding walk in coin space}

The limit theorem for a walk self-avoiding in coin space as presented in the manuscript is given by
\begin{equation}
 \lim_{t\to\infty}P\left(\frac{X_t}{t}\leq x,\,\frac{Y_t}{t}\leq y\right)
 =\int_{-\infty}^x du\,\int_{-\infty}^y dv\,f_{sc}(u,v)\left\{g_1(u,v)+g_2(u,v)\right\}
 \times\eta_{sc}(u,v;\alpha,\beta,\gamma,\delta)I_{\mathcal{D}_{sc}}(u,v),
\end{equation}
where
\begin{align}
 f_{sc}(x,y)=&\frac{1}{\pi^2(1-4x^2)(1-4y^2)\sqrt{D_{sc}(x,y)}},\\
 D_{sc}(x,y)=&81(x^4+y^4)-18x^2y^2-18(x^2+y^2)+1,\\
 g_1(x,y)=&\Bigl\{648(x^4+y^4)+576x^2y^2-324(x^2+y^2)
 +53+4\left\{7-18(x^2+y^2)\right\}\sqrt{D_{sc}(x,y)}\Bigr\}^{\frac{1}{2}},\\
 g_2(x,y)=&\Bigl\{648(x^4+y^4)+576x^2y^2-324(x^2+y^2)
 +53-4\left\{7-18(x^2+y^2)\right\}\sqrt{D_{sc}(x,y)}\Bigr\}^{\frac{1}{2}},\\
 \eta_{sc}(x,y;\alpha,\beta,\gamma,\delta)
 =&1+2x\Bigl[|\alpha|^2-|\delta|^2
     \quad+\Re\left\{(-\alpha+\gamma-\delta)\overline{\beta}+(\alpha+\beta-\delta)\overline{\gamma}\right\}\Bigr]\nonumber\\
 &\quad-2y\Bigl[|\beta|^2-|\gamma|^2
  +\Re\left\{(\beta+\gamma+\delta)\overline{\alpha}+(\alpha-\beta+\gamma)\overline{\delta}\right\}\Bigr],\\
  I_{\mathcal{D}_{sc}}(x,y)=&\left\{\begin{array}{ll}
			 1&\left(D_{sc}(|x|,|y|)>0,\,0\leq |x|,|y|\leq \frac{1}{3}\right),\\[1mm]
			       0&(\mbox{otherwise}).\end{array}\right.
\end{align}

\begin{proof}
Using Fourier analysis~\cite{GJS04} the state of the self-avoiding walker at time $t$ can be written as 
\begin{equation}
   \miniket{\hat\Psi_t(k_x,k_y)}=\sum_{(x,y)\in\mathbb{Z}^2}e^{-i(k_x x+k_y y)}\miniket{\psi_t(x,y)}.
   \label{fourier}
\end{equation}
The time evolution in Fourier space is given by
\begin{equation}
   \miniket{\hat\Psi_{t+1}(k_x,k_y)}=\hat{C}^{sc}(k_x,k_y)\miniket{\hat\Psi_t(k_x,k_y)},
\end{equation}
where $\hat{C}^{sc}(k_x,k_y)=\hat{R}(k_x,k_y)C^{sc}$ and
\begin{equation}
  \hat{R}(k_x,k_y)=
  \begin{pmatrix}
    e^{ik_x}&0&0&0\\
    0&e^{-ik_y}&0&0\\
    0&0&e^{ik_y}&0\\
    0&0&0&e^{-ik_x}
  \end{pmatrix}\;.
\end{equation}
Therefore, we get
\begin{equation}
\miniket{\hat\Psi_t(k_x,k_y)}=\hat{C}^{sc}(k_x,k_y)^t\miniket{\hat\Psi_0(k_x,k_y)}.
\label{hatPsit}
\end{equation}

Using the eigenvalues $\lambda_j(k_x,k_y)$ and the normalized eigenvectors $\miniket{v_j(k_x,k_y)}\,(j=1,2,3,4)$ of the matrix $\hat{C}^{sc}(k_x,k_y)$, the $(r_1, r_2)$-th joint moments ($r_1, r_2=0,1,2,\ldots$) of $(X_t,Y_t)$ can be expressed as
\begin{align}
  \mathbb{E}(X_t^{r_1}Y_t^{r_2})=&\sum_{(x,y)\in\mathbb{Z}^2} x^{r_1}y^{r_2}\mathbb{P}[(X_t,Y_t)=(x,y)]\\
  =&\int_{-\pi}^\pi\frac{dk_x}{2\pi}\int_{-\pi}^\pi\frac{dk_y}{2\pi}\sand{\hat\Psi_t(k_x,k_y)}{D_x^{r_1}D_y^{r_2}}{\hat\Psi_t(k_x,k_y)}\\
  =&(t)_{r_1+r_2}\int_{-\pi}^\pi\frac{dk_x}{2\pi}\int_{-\pi}^\pi\frac{dk_y}{2\pi}\sum_{j=1}^4\left\{\frac{D_x\lambda_j(k_x,k_y)}  {\lambda_j(k_x,k_y)}\right\}^{r_1}
  \left\{\frac{D_y\lambda_j(k_x,k_y)}{\lambda_j(k_x,k_y)}\right\}^{r_2}|\miniprod{v_j(k_x,k_y)}{\hat\Psi_0(k_x,k_y)}|^2\nonumber\\
  &+O(t^{r_1+r_2-1})
\end{align}
with $D_x=i(\partial{}/\partial{k_x})$, $D_y=i(\partial{}/\partial{k_y})$ and $(t)_r=t(t-1)\times\cdot\cdot\cdot\times(t-r+1)$, where $\mathbb{E}(X)$ denotes the expected value of $X$.
For the joint moments of the rescaled walker's position $(X_t/t, Y_t/t)$, by setting $D_x\lambda_j(k_x,k_y)/\lambda_j(k_x,k_y)=x, D_y\lambda_j(k_x,k_y)/\lambda_j(k_x,k_y)=y$ after $t\to\infty$, we get the convergence theorem
\begin{align}
\lim_{t\rightarrow\infty}\mathbb{E}\left[\left(\frac{X_t}{t}\right)^{r_1}\left(\frac{Y_t}{t}\right)^{r_2}\right]
=&\int_{-\pi}^\pi\frac{dk_x}{2\pi}\int_{-\pi}^\pi\frac{dk_y}{2\pi}\sum_{j=1}^4\left\{\frac{D_x\lambda_j(k_x,k_y)}{\lambda_j(k_x,k_y)}\right\}^{r_1}
\left\{\frac{D_y\lambda_j(k_x,k_y)}{\lambda_j(k_x,k_y)}\right\}^{r_2}|\miniprod{v_j(k_x,k_y)}{\hat\Psi_0(k_x,k_y)}|^2\\
=&\int_{-\infty}^\infty dx\int_{-\infty}^\infty dy\,x^{r_1}y^{r_2} f_{sc}(x,y)\left\{g_1(x,y)+g_2(x,y)\right\}\eta_{sc}(x,y;\alpha,\beta,\gamma,\delta)I_{\mathcal{D}_{sc}}(x,y).
\label{limE}
\end{align}
Equation~\eqref{limE} guarantees Theorem~1.

\hfill\qed

\end{proof}




\end{widetext}
\end{document}